\documentclass[pra,aps,amssymb,twocolumn,noshowpacs]{revtex4-1}
\usepackage{color}
\usepackage{graphicx}
\usepackage{hyperref}
\usepackage{amsmath}
\usepackage{latexsym}
\usepackage{amssymb}
\usepackage{mathrsfs}
\usepackage{mathtools}
\usepackage{layout}
\usepackage{verbatim}
\usepackage{bm}
\usepackage{amsfonts,epsfig}

\newcommand\startsupplement{
    \makeatletter
       \setcounter{table}{0}
       \renewcommand{\thetable}{S\arabic{table}}
       \setcounter{figure}{0}
       \renewcommand{\thefigure}{S\arabic{figure}}
       \setcounter{equation}{0}
       \renewcommand{\theequation}{S\arabic{equation}}
    \makeatother}

\begin{document}
\title{Deutsch, Toffoli, and CNOT Gates via Rydberg Blockade of Neutral Atoms}
\date{\today}
\author{Xiao-Feng Shi}
\affiliation{School of Physics and Optoelectronic Engineering, Xidian University, Xi'an 710071, China}

\begin{abstract}
Universal quantum gates and quantum error correction~(QEC) lie in the heart of quantum information science. Large-scale quantum computing depends on a universal set of quantum gates, in which some gates may be easily carried out, while others are hard with a certain physical system. There is a unique three-qubit quantum gate called the Deutsch gate~[$\mathbb{D}(\theta)$], from which alone a circuit can be constructed so that any feasible quantum computing is attainable. As far as we know, however, $\mathbb{D}(\theta)$ has not been demonstrated. Here we design an easily realizable $\mathbb{D}(\theta)$ by using Rydberg blockade of neutral atoms, where $\theta$ can be tuned to any value in $[0,\pi]$ by adjusting the strengths of external control fields. Using similar protocols, we further show that both the Toffoli and CNOT gates can be achieved with only three laser pulses. The Toffoli gate, being universal for classical reversible computing, is also useful for QEC that plays an important role in quantum communication and fault-tolerant quantum computation. The possibility and briefness to realize these gates shed new light on the study of quantum information with neutral atoms.

\end{abstract}
\maketitle

\section{INTRODUCTION}
Quantum computation and communication lie in the heart of applicability of quantum mechanics, where the former can facilitate fascinating computing tasks beyond the capability of classical computers, and the latter can realize exceedingly secure communication~\cite{Nielsen2000}. The design of large-scale quantum computers depends on availability of at least one universal set of quantum gates that can represent any unitary operation~\cite{Deutsch1989}. A well-known universal set of quantum gates 
consists of three or four single-qubit gates plus the two-qubit controlled-NOT~(CNOT) gate~\cite{Williams2011}, which has been targeted for decades~\cite{PhysRevLett.75.4714} by experimentalists with, e.g., single photons~\cite{Pryde2003}, electrons in silicons~\cite{Veldhorst2015,Zajac2017}, superconducting circuits~\cite{Yamamoto2010,PhysRevLett.109.060501,Barends2014}, atomic ions~\cite{Ballance2016}, and neutral Rydberg atoms~\cite{Isenhower2010,Maller2015}. While certain quantum gates could be prepared very accurately~\cite{Barends2014,Ballance2016,Gaebler2016,Xia2015,Wang2016}, to make ready a complete universal set of quantum gates in a scalable physical platform is challenging.

\begin{figure}
\includegraphics[width=3.3in]
{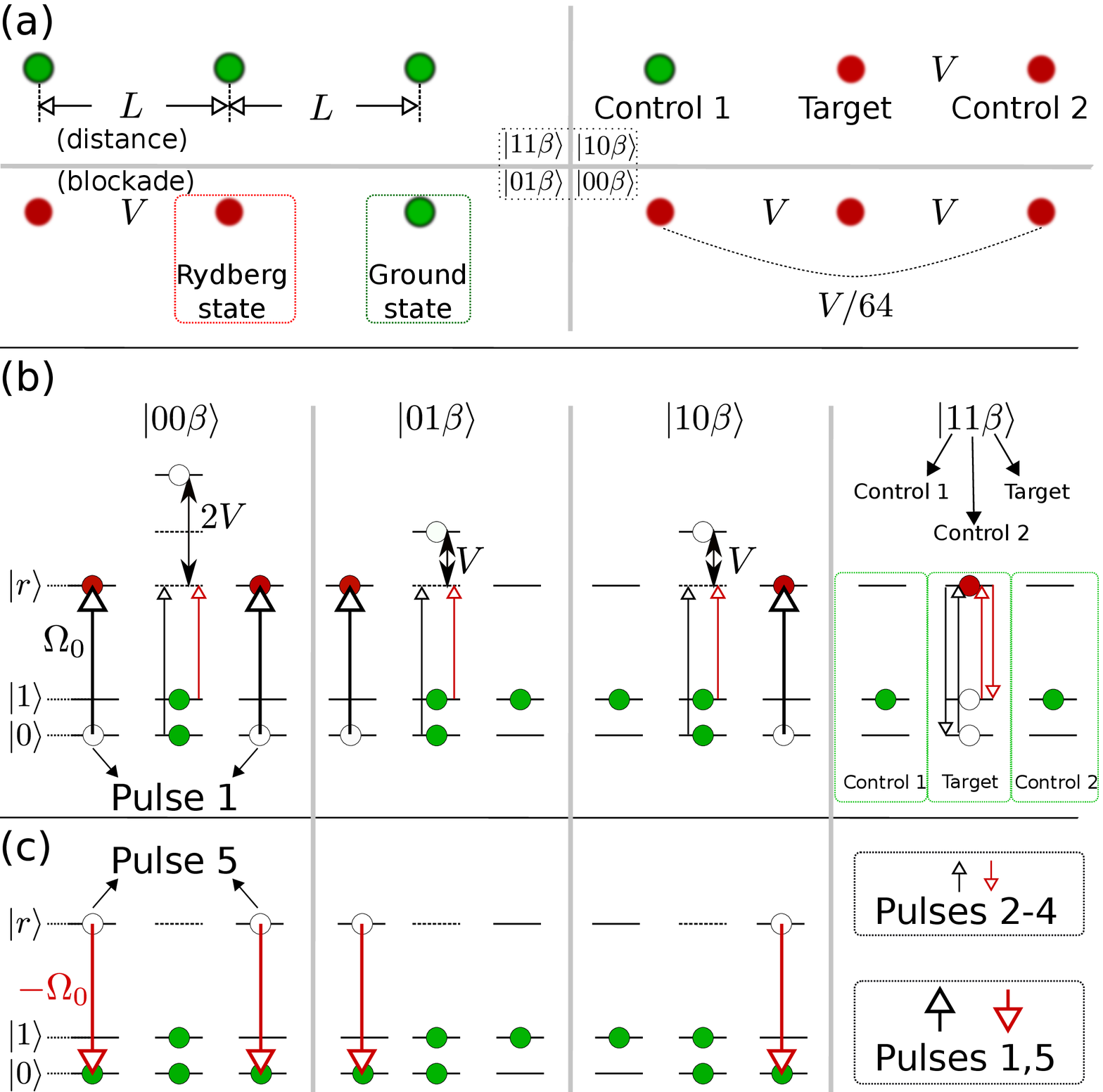}
 \caption{Schematic of the Deutsch gate protocol. (a) shows the three-qubit configuration and their blockade interaction for four general input states. In the cartoon, the two outermost atoms serve as control qubits, and the central one as target. Blockade interaction happens if $\alpha_1\alpha_2=0$ for the input state $|\alpha_1\alpha_2\beta\rangle$, i.e., when at least one of the two control qubits are initialized in $|0\rangle$. When both control qubits are in Rydberg state $|r\rangle$, there is a residue interaction $V/2^6$ between them. (b) and (c) illustrate the five-pulse protocol, where large~(or, small) arrows indicate pulses 1 and 5~(or, 2-4). In (c), the Rabi frequency for pulse-5 is $-\Omega_0$, so that the three general input states $|00\beta\rangle,|01\beta\rangle$, and $|10\beta\rangle$ stay intact when the gate completes.\label{figure01} }
\end{figure}

Another route toward realizing quantum computing is to focus on the realization of only one type of quantum gate. As proposed by Deutsch~\cite{Deutsch1989}, reliable large-scale quantum computing is possible based on the following three-qubit quantum gate
\begin{eqnarray}
 \mathbb{D}(\theta) &=& \left(\begin{array}{cc}
    \openone_{6} &0  \\
   0& D_0(\theta)
 \end{array}
  \right),\label{DGate}
\end{eqnarray}
written in the basis states $\{|000\rangle, |001\rangle,|010\rangle,\cdots,|111\rangle \}$, with $|0(1)\rangle$ being a qubit state, where $\openone_{6}$ is the $6\times6$ identity matrix, and 
\begin{eqnarray}
 D_0 (\theta)&=& \left(\begin{array}{cc}
   i\cos\theta &\sin\theta \\
   \sin\theta& i\cos\theta
  \end{array}
  \right),\label{DGate01}
\end{eqnarray}
where $2\theta/\pi$ is a fixed irrational number. At a glance, the realization of $\mathbb{D}(\theta)$ not only requires to couple three qubits but also needs quite special form of the coupling among them. Such requirement is daunting concerning real physical systems. To the best of our knowledge, there has been no theoretical proposal to realize the Deutsch gate directly, not to speak of its experimental realization.

In this Letter, we introduce a protocol to realize $\mathbb{D}(\theta)$ with neutral atoms, where $\theta$ is tunable in $[0,\pi]$ by adjustment of laser field strengths. By using dipole-dipole interaction between nearby atoms in Rydberg states~\cite{PhysRevLett.85.2208,Saffman2016}, we show that five pulses of lasers are sufficient to complete the sequence of the Deutsch gate. The Rydberg blockade used in our method has been verified by numerous experiments~\cite{Saffman2016}, thus our gate protocol can be easily realized. This makes it possible to build quantum computers based on a single type of gate.

Our Deutsch gate protocol can be extended to realize the Toffoli gate, i.e., the three-qubit controlled-controlled-NOT gate, which is universal for classical reversal computing~\cite{Williams2011}. More important, the Toffoli gate is useful in quantum error correction, which is relevant to quantum state initialization, manipulation, and measurement in general, i.e., valuable in quantum communication as well as in fault-tolerant quantum computation. For this reason, the Toffoli gate was experimentally studied in several different physical platforms~\cite{Cory1998,Lanyon2009,Monz2009,Reed2012,Fedorov2012}, but has never been demonstrated with neutral atoms, as far as we know. Remarkably, our neutral-atom Toffoli gate can be accomplished with only three laser pulses, being much simpler than to simulate it with several single- and two-qubit gates or via five two-qubit gates~\cite{Yu2013}. Along the way, we find a new CNOT gate protocol that can be accomplished with three laser pulses, which compares favorably to a traditional Rydberg CNOT gate in, e.g., Refs.~\cite{Isenhower2010},~\cite{Zhang2010}, and~\cite{Maller2015} that respectively require seven optical pulses, five optical pulses, and a combination of three optical and two microwave pulses. The introduction of the Deutsch and Toffoli gates, and the new method to achieve the CNOT gate via Rydberg blockade brings new opportunity to study quantum information with neutral atoms~\cite{PhysRevLett.85.2208,Saffman2010,Saffman2016}.

\section{A three-qubit system}
Our Deutsch gate operates in a three-qubit system as illustrated in Fig.~\ref{figure01}(a), where three neutral $^{133}$Cs atoms are trapped along the quantization axis before the gate sequence, with the distance between two nearest atoms equal to $L$. The two outermost atoms serve as control qubits~(labeled as control 1 and 2), while the center one as target qubit. This spatial arrangement of the three qubits shall not be confused with the way we write the basis state $|\alpha_1\alpha_2\beta\rangle$ of Eq.~(\ref{DGate}),  where the first two numbers $\alpha_1$ and $\alpha_2$ in the ket denote states of the two control qubits, and the last number $\beta$ denotes state of the target qubit, where ${\alpha_k,\beta}\in\{0,1\}$ with $k=1$ or $2$.  As for the qubit states, one can choose from the hyperfine ground states \begin{eqnarray}
  |0(1)\rangle &=& |6s_{1/2},F=3(4),m_F=0\rangle \nonumber
\end{eqnarray}
for both the control and target qubits~\cite{Maller2015}.

We consider laser pulses that excite the qubit state $|\alpha\rangle$ to a Rydberg state $|r\rangle$, where $|\alpha\rangle$ can be the qubit state $|0\rangle$ of control 1 or 2, or one of the states $|0\rangle$ and $|1\rangle$ of the target qubit, as shown in Figs.~\ref{figure01}(b) and~\ref{figure01}(c). In this case, when the target qubit and one of the two control qubits are in Rydberg state, there arises an interaction $V=C_6/L^6$, where $C_6$ is the van der Waals interaction~(vdWI) coefficient~\cite{Gallagh2005}. If all three qubits are in Rydberg state, the total interaction is $2V$ plus a small interaction $V/2^6$ between the two control qubits, as shown in Fig.~\ref{figure01}. When resonant laser fields with Rabi frequencies much smaller than $V$ are used upon the target, it is hard to excite the target qubit to the Rydberg state $|r\rangle$ if one of the two control qubits is already in state $|r\rangle$. This blockade mechanism is the essence of our protocols.

\begin{figure}
\includegraphics[width=3.3in]
{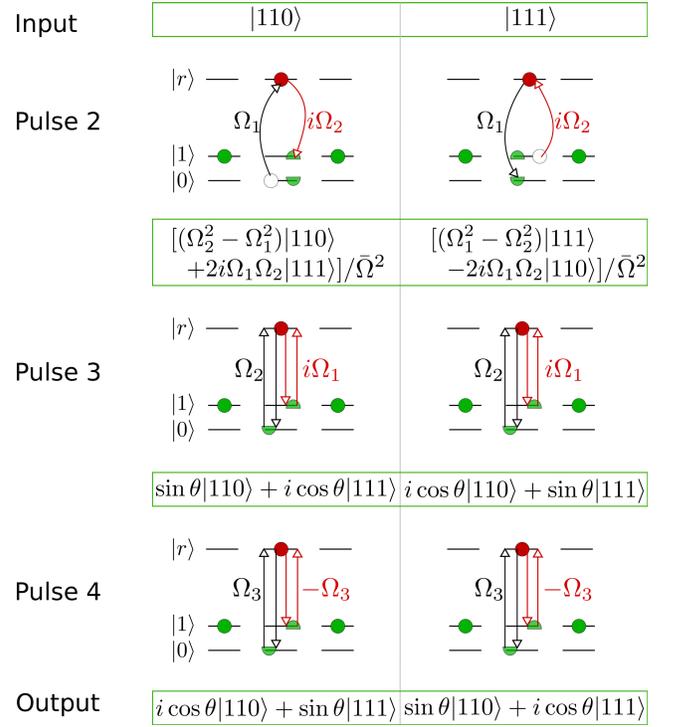}
 \caption{Pulses 2-4 of the Deutsch gate. Action of pulses 2-4 on the two input states $|110\rangle$ and $|111\rangle$ leads to the transform of Eq.~(\ref{DGate01}). \label{figure02} }
\end{figure}

\begin{figure}
\includegraphics[width=2.5in]
{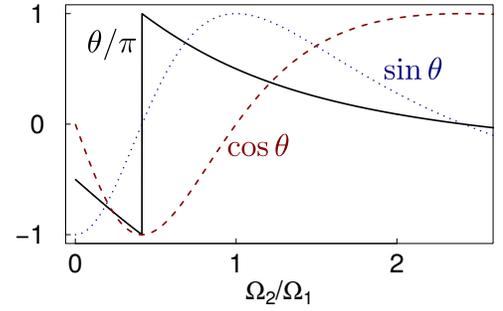}
 \caption{Tunability of the angle $\theta$ in the Deutsch gate. $\theta$ is smoothly tunable in $[\pi,0]$ when $\Omega_2/\Omega_1\in[\sqrt{3-2\sqrt2},\sqrt{3+2\sqrt2}]$. \label{figure03} }
\end{figure}
\section{ A five-pulse protocol of $\mathbb{D}(\theta)$}
We proceed to show a five-pulse sequence for $\mathbb{D}(\theta)$, based on which we will briefly explain a three-pulse protocol for the Toffoli gate. For the sake of convenience, we use pulse-$k$ to denote the $k$th pulse, where $k=1-5$. 

Pulses-1 and 5 are illustrated by the larger arrows in Figs.~\ref{figure01}(b) and~\ref{figure01}(c), respectively, while pulses 2-4 are indicated by the smaller arrows in Fig.~\ref{figure01}(b). Because pulses 2-4 are essential to induce the transform of Eq.~(\ref{DGate01}), they are highlighted in Fig.~\ref{figure02}. The gate sequence starts with a $\pi$ pulse of Rabi frequency $\Omega_0$ upon the two $|0\rangle$ states of the two control qubits, so as to induce the following map,
\begin{eqnarray}
 && \{|000\rangle,|001\rangle,|010\rangle, |011\rangle,|100\rangle,|101\rangle,|110\rangle, |111\rangle\}\mapsto\nonumber\\ &&-\{ |rr0\rangle, |rr1\rangle, i|r10\rangle,  i|r11\rangle, i|1r0\rangle, i|1r1\rangle,-|110\rangle, -|111\rangle\}.\nonumber
\end{eqnarray}
The process $\{|000\rangle,|001\rangle\}\mapsto-\{ |rr0\rangle, |rr1\rangle\}$ above is subject to a residue blockade effect which can be minimized by increasing the ratio $|64\Omega_0/V|$. 

Pulse-2 is a $2\pi$ pulse for the transition chain 
\begin{eqnarray}
|0\rangle_{t}\xleftrightarrow  [\text{duration:~}t=2\pi/\bar{\Omega}]  { \Omega_1} |r\rangle_{t}\xleftrightarrow  [\text{duration:~}t=2\pi/\bar{\Omega}]  {i\Omega_2 }    |1\rangle_{t}, \label{pulse2}
\end{eqnarray}
where $\bar{\Omega} = \sqrt{\Omega_1^2+\Omega_2^2}$, the subscript `t' denotes target qubit, and the process $|r\rangle_{t}\xleftrightarrow  {i\Omega_2 }    |1\rangle_{t}$ above represents the dynamics governed by $i\Omega_2[ |r\rangle \langle 1|-\text{H.c.}]/2$. When $\bar\Omega\ll V$, the six states $-\{|rr0\rangle, |rr1\rangle, i|r10\rangle,  i|r11\rangle, i|1r0\rangle, i|1r1\rangle\}$ stay intact due to the blockade effect~(an extra dynamical phase for the first two of the six states can be avoided~\cite{supple}), while the other two states $\{|110\rangle, |111\rangle\}$ form a $\Lambda$-type three-level system with $|11r\rangle$. As can be easily verified, the input states $|110\rangle$ and $|111\rangle$ experience the following state evolutions as a result of pulse-2,
\begin{eqnarray}
  |110\rangle& \rightarrow&|\psi_0(t)\rangle= [(\Omega_2^2-\Omega_1^2)|110\rangle +2i\Omega_1\Omega_2|111\rangle]/\bar{\Omega}^2,\nonumber\\
  |111\rangle &\rightarrow &|\psi_1(t)\rangle=[(\Omega_1^2-\Omega_2^2)|111\rangle -2i\Omega_1\Omega_2|110\rangle]/\bar{\Omega}^2, \nonumber\\
  \label{pulse2-1}
\end{eqnarray}
where we assume the time is $0$ at the beginning of pulse-2.

Pulse-3 is also a $2\pi$ pulse upon the target qubit. In reference to pulse-2, the magnitudes of the two Rabi frequencies for the two transitions in pulse-3 are swapped compared with those in Eq.~(\ref{pulse2}), i.e., in pulse-3 we have
\begin{eqnarray}
|0\rangle_{t}\xleftrightarrow  [2\pi/\bar{\Omega}]  { \Omega_2} |r\rangle_{t}\xleftrightarrow  [2\pi/\bar{\Omega}]  {i\Omega_1 }    |1\rangle_{t}, \nonumber\label{pulse3}
\end{eqnarray}
and similar to Eq.~(\ref{pulse2-1}), one can show that
\begin{eqnarray}
  |\psi_0(t)\rangle& \rightarrow&|\psi_0(2t)\rangle=\sin\theta |110\rangle +i\cos\theta|111\rangle,\nonumber\\
|\psi_1(t)\rangle&\rightarrow &|\psi_1(2t)\rangle=i\cos\theta |110\rangle +\sin\theta|111\rangle,
  \label{pulse3-1}
\end{eqnarray}
where
\begin{eqnarray}
  \sin\theta &=& [6\Omega_1^2\Omega_2^2-(\Omega_1^4+\Omega_2^4)]/\bar{\Omega}^4,\nonumber\\
  \cos\theta &=& 4\Omega_1\Omega_2(\Omega_2^2 - \Omega_1^2) /\bar{\Omega}^4.\label{angle}
\end{eqnarray}

Pulse-4 is also a $2\pi$ pulse upon the target qubit. Different from pulse-2 and pulse-3, here we have two Rabi frequencies of equal magnitude $\Omega_3$ but with a $\pi$ phase difference for the two transitions in the following chain,
\begin{eqnarray}
|0\rangle_{t}\xleftrightarrow  [t'=\sqrt2\pi/\Omega_3]  { \Omega_3} |r\rangle_{t}\xleftrightarrow  [t'=\sqrt2\pi/\Omega_3]   {-\Omega_3 }    |1\rangle_{t}, \label{pulse4}
\end{eqnarray}
so that
\begin{eqnarray}
  |\psi_0(2t)\rangle& \rightarrow&|\psi_0(2t+t')\rangle=i\cos\theta |110\rangle +\sin\theta|111\rangle, \nonumber\\
|\psi_1(2t)\rangle&\rightarrow &|\psi_1(2t+t')\rangle=\sin\theta |110\rangle +i\cos\theta|111\rangle,
  \label{pulse4-1}
\end{eqnarray}
which completes the transform of Eq.~(\ref{DGate01}). Pulse-5 is a $\pi$ pulse upon the two control qubits, as shown in Fig.~\ref{figure01}(c), resulting in the following map  
\begin{eqnarray}
 &&-\{ |rr0\rangle, |rr1\rangle, i|r10\rangle,  i|r11\rangle, i|1r0\rangle, i|1r1\rangle\} \nonumber\\ && \mapsto\{|000\rangle,|001\rangle,|010\rangle, |011\rangle,|100\rangle,|101\rangle\},\nonumber
\end{eqnarray}
which, together with Eqs.~(\ref{pulse2-1}),~(\ref{pulse3-1}), and~(\ref{pulse4-1}), constitute the Deutsch gate protocol in Eq.~(\ref{DGate}). The angle $\theta$ in Eq.~(\ref{angle}) can be tuned to any value in $[0,\pi]$ by adjustment of the ratio between $\Omega_1$ and $\Omega_2$, as shown in Fig.~\ref{figure03}. For any fixed irrational $2\theta/\pi$, $\mathbb{D}(\theta)$ is universal for quantum computing~\cite{Deutsch1989}.

When pulses-2 and 3~(see Fig.~\ref{figure02}) are excluded in the above protocol, one can easily verify that a sequence of pulses-1, 4, and 5 will give rise to the Toffoli gate $\mathbb{D}(\pi/2)$. Alternatively, the five-pulse protocol above can also realize the Toffoli gate as long as $\Omega_1=\Omega_2$ in Eq.~(\ref{angle}), although being a little tedious compared with a three-pulse sequence. The Toffoli gate, being universal for classical reversal computing~\cite{Williams2011} and useful for quantum error correction~\cite{Cory1998,Reed2012}, was experimentally demonstrated with, e.g., nuclear spins~\cite{Cory1998}, photons~\cite{Lanyon2009}, trapped ions~\cite{Monz2009} and superconducting circuits~\cite{Reed2012,Fedorov2012}.

\section{Gate performance}
Here we show the performance quality of the Deutsch gate by taking the Rydberg state $|r\rangle=|84p_{3/2},m_J=3/2\rangle$ of $^{133}$Cs as an example, whose blockade mechanism was experimentally demonstrated in~\cite{Hankin2014}. The state $|r\rangle$ has a lifetime of $\tau=1.59$~ms~(313~$\mu$s) in a temperature of 4.2~(300)~K~\cite{Beterov2009} and a vdWI coefficient of $C_6/2\pi=-633$~GHz$\mu m^6$~(calculated~\cite{Shi2014} using quantum defects in~\cite{Lorenzen1984,PhysRevA.35.4650}, see~\cite{supple}). The minimal time to implement the five-pulse protocol of $\mathbb{D}(\theta)$ is 
\begin{eqnarray}
 T_g &=& 2\pi\left( \frac{1}{\Omega_0}+ \frac{2}{\bar\Omega}+ \frac{1}{\sqrt2\Omega_3}  \right).\label{gatetime}
\end{eqnarray}

\begin{figure}
\includegraphics[width=3.1in]
{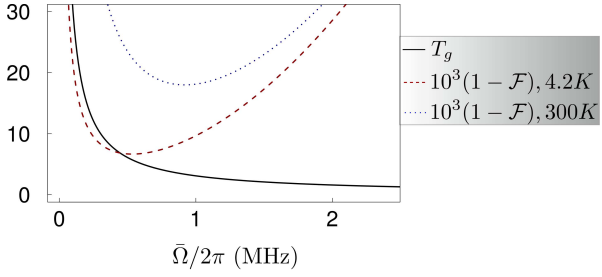}
 \caption{Performance quality of the Deutsch gate. Plotted here are the gate time $T_g$~(in units of $\mu$s) and rescaled fidelity error in an environment of $4.2~(300)$~K as a function of $\bar\Omega$ when $\theta=\sin^{-1}(7/25)$~(realized with $\Omega_2/\Omega_1=2$) in $\mathbb{D}(\theta)$. Parameters can be found around Eq.~(\ref{gatetime}). \label{figure04} }
\end{figure}

Three mechanisms result in imperfect gate operation: the decay of Rydberg state, the blockade interaction $V/64$ which hampers the transition $|00\beta\rangle\leftrightarrow|rr\beta\rangle$, and population leakage due to six two-photon transitions such as $|rr0\rangle\leftrightarrow|rr1\rangle$ formed via the highly detuned level $|rrr\rangle$~\cite{supple}. Due to the requirement $|V|/64\ll \Omega_0$ and $ \Omega_k\ll |V|$ with $k=1-3$~[see texts around Eq.~(\ref{pulse2})], we choose, as an example, $L=6\mu$m, $\Omega_0/2\pi=10$~MHz, $\Omega_2/\Omega_1=2$, and $\bar\Omega=\sqrt2\Omega_3 <2\pi\times2.3$~MHz to show the rescaled fidelity error in Fig.~\ref{figure04}. With a temperature of $4.2~(300)$~K, we find a minimal error $6.7~(18)\times10^{-3}$ when $\bar\Omega/2\pi=0.54~(0.92)$~MHz. A major contribution to the gate error in Fig.~\ref{figure04} is decay of Rydberg state, which can be minimized by using higher Rydberg states and larger $\Omega_0$ that respectively give lower decay rates and shorter gate times. However, since our protocols are based on the traditional blockade mechanism, their fidelity errors should be over $10^{-3}$ as in a traditional Rydberg blockade gate~\cite{Zhang2012}. But in principle, the fidelity error can be significantly suppressed by extending our protocols to those using detuned Rabi transitions, where gate operation times can be in the $0.1$-$\mu$s regime or even shorter~\cite{Shi2017}.

\begin{figure}
\includegraphics[width=3.4in]
{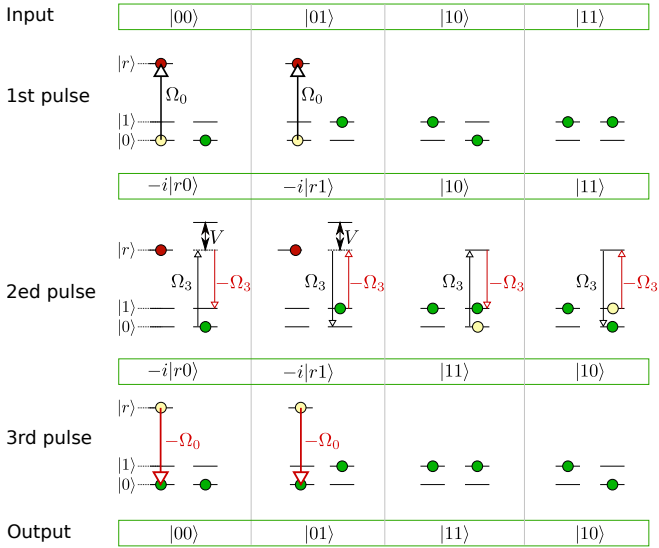}
 \caption{State evolution in a three-pulse CNOT gate. Uninvolved Rydberg states are hidden for the sake of clarity. \label{figure05} }
\end{figure}

\section{A three-pulse protocol of the CNOT gate}
A straightforward extension of the Deutsch gate protocol is to realize the CNOT gate by three laser pulses. With a two-qubit system of the control 1 and the target qubit in Fig.~\ref{figure01}(a), the CNOT gate sequence is illustrated in Fig.~\ref{figure05}. When the control qubit is initialized in state $|0\rangle$, it is excited to the state $|r\rangle$ during the first pulse. Then the second pulse with Rabi frequency $|\Omega_3|\ll |V|$ will leave the two states $|r0\rangle$ and $|r1\rangle$ intact. However, the three states $|10\rangle$, $|11\rangle$, and $|1r\rangle$ form a $\Lambda$-type three-level system in the second pulse, and similar to Eq.~(\ref{pulse4}), there is also a $\pi$ phase difference for the two Rabi frequencies in the transition chain $|10\rangle\leftrightarrow |1r\rangle \leftrightarrow |11\rangle$. Then, it is easily verified that the two input states $|10\rangle$ and $|11\rangle$ evolve to $|11\rangle$ and $|10\rangle$, respectively, when we choose a pulse duration of $t'=\sqrt2\pi/\Omega_3$ for the second pulse. After the third pulse that has a Rabi frequency of equal magnitude to that in the first pulse, but with a $\pi$ phase difference, the states $|r0\rangle$ and $|r1\rangle$ return to $|00\rangle$ and $|01\rangle$, respectively, completing the CNOT gate. Here, the central physics is that the control qubit, once in Rydberg state, prevents the state swap between $|0\rangle$ and $|1\rangle$ of the target qubit, analogous to the parallelized CNOT gate proposed in~\cite{Muller2009} via an interaction-induced breakdown of adiabatic state transfer in the configuration of electromagnetically induced transparency.

Compared to the CNOT gate in~\cite{Maller2015}, the more laser fields in our CNOT gate may result in larger laser-noise-induced gate error. For lasers, however, the frequency and phase fluctuations can be negligibly small~\cite{Wineland1998,Saffman2005}, and the intensity noise can also be less than $0.1\%$, as demonstrated in a recent experiment with Rydberg states of alkali-metal atoms~\cite{Helmrich2017}, thus it is possible to achieve high fidelity with our method.

\section{ Discussion and conclusions}
Our Deutsch, Toffoli, and CNOT protocols are derived from the Rydberg blockade mechanism, a method used in recent logic gate experiments with Rydberg atoms~\cite{Isenhower2010,Wilk2010,Zhang2010,Maller2015,Jau2015}, except that we require different relative phase shifts $\pi/2$ or $\pi$ between different laser Rabi frequencies, as shown by the colored arrows in Figs.~\ref{figure01}, \ref{figure02}, and~\ref{figure05}. This latter requirement, frequently assumed in theoretical works on Rydberg quantum gates~\cite{Zhang2012,Beterov2016}, is achievable by using linear optics technique~\cite{Kok2007} since the phase of a Rabi frequency is determined by the laser fields and constant atomic dipole matrix elements. This means that our gate protocols can be easily verified in experiments.

Since it is challenging to prepare every quantum gate in a universal set with high accuracy, our method offers a more favorable route to build a quantum computer by designing only one type of gate such as $\mathbb{D}(\theta)$, compared with the strategy of designing several different types of gates. Moreover, our protocol allows realization of $\mathbb{D}(\theta)$ with any $\theta$, thus offers an over-complete family of universal gates that should aid the construction of quantum computers~\cite{Williams2011}.

The Toffoli gate can be used for quantum error correction~\cite{Cory1998,Lanyon2009,Monz2009,Reed2012,Fedorov2012}, which is a useful technique in preparation, manipulation, and measurement of quantum states especially because most quantum systems that can be used as quantum information carrier are vulnerable to noise. Thus our Toffoli gate protocol can serve in both quantum computation and communication.

In summary, we have proposed a five-pulse protocol to realize the Deutsch gate $\mathbb{D}(\theta)$ with neutral atoms, where $\theta$ is tunable by adjusting the strengths of laser fields. We further show that both the Toffoli and CNOT gates can be implemented in a similar manner, each with only three laser pulses. Given the applicability of the Deutsch, CNOT, and Toffoli gates in universal quantum computing, classical reversal computing, and quantum error correction, the briefness to implement these gates may help to advance quantum computation and communication science.

\section*{ACKNOWLEDGMENTS}
The author thanks Yan Lu for fruitful discussions and acknowledges support from the Fundamental Research Funds for the Central Universities and the 111 Project (B17035).

%
 
\newpage
~~
\newpage
\onecolumngrid
\startsupplement
\begin{center}
  \large{\bf Supplemental Material for ``Deutsch, Toffoli, and CNOT gates via Rydberg blockade of neutral atoms''}
\end{center}
\author{Xiao-Feng Shi}
\affiliation{School of Physics and Optoelectronic Engineering, Xidian University, Xi'an 710071, China}

\twocolumngrid

In this Supplemental Material, we provide additional information about the fidelity error of the Deutsch gate protocol. We also show how to avoid the unwanted phase accumulation in the input states $|000\rangle$ and $|001\rangle$ because of the residue interaction $V/64$ between the two control qubits, shown in Fig. 1(a) of the main text. The following discussions apply to both the Deutsch and the Toffoli gate protocols.

Before the analysis of the gate fidelity error, we briefly introduce the configuration of atomic levels for the gate example in Fig. 4, which is essential for an adequate modeling of the three-qubit Deutsch or Toffoli gate. We consider using $^{133}$Cs atoms as qubits, with the qubit states
\begin{eqnarray}
  |0(1)\rangle &=& |6s_{1/2},F=3(4),m_F=0\rangle \nonumber
\end{eqnarray}
for both the control and target qubits, as in the experiments of~\cite{Maller2015}. By applying right-hand circularly polarized laser fields, both qubit states can be excited to a combination of $|np_{3/2},m_J=3/2,m_I=-1/2\rangle$ and $|np_{1/2},m_J=1/2,m_I=1/2\rangle$ according to the dipole selection rule. However, because the oscillator strength between $|0(1)\rangle$ and $|np_{1/2},m_J=1/2,m_I=1/2\rangle$ is extremely small for a large principal quantum number $n$~\cite{Hankin2014}, we can neglect it, thus $|r\rangle=|np_{3/2},m_J=3/2,m_I=-1/2\rangle$ is excited. We choose the Rydberg state of principal quantum number $n=84$ because its direct excitation has been demonstrated in laboratories~\cite{Hankin2014}. By using the data of quantum defects in Refs.~\cite{Lorenzen1984,PhysRevA.35.4650}, we calculate~\cite{Shi2014} the vdWI coefficient, $C_6/2\pi=-633$~GHz$\mu m^6$. The value of $C_6$ in experiments, however, can deviate from the calculated value when atomic energy gaps change in response to external fields.

\section{Remove phase accumulation for $|000\rangle$ and $|001\rangle$ }
During pulses 2-4 and for the input states $|000\rangle$ and $|001\rangle$, there is an interaction $V/64$ for the states $|rr0\rangle$ and $|rr1\rangle$. This will cause a phase accumulation $\varphi=-T_{\overline{234}}V/64$, where the duration from the beginning of pulse-2 to that of pulse-5 is $T_{\overline{234}}=2\pi\left( \frac{2}{\bar\Omega}+ \frac{1}{\sqrt2\Omega_3}  \right)$, which is quite large for small $\bar\Omega$ in Fig.~4, and decreases to about $0.5\pi$ on the right edge of Fig.~4. At least two methods can be employed to remove this unwanted phase accumulation, as shown below.

First, one can choose parameters so that $\varphi$ becomes $2N\pi$, where $N$ is an integer. Because $e^{2iN\pi}=1$, the phase term $\varphi$ is effectively removed in this case. Take the gate fidelity at 4.2~K shown in Fig.~4 for example; the minimal gate error $6.7\times10^{-3}$ occurs when $\bar{\Omega}/2\pi=0.54$~MHz, around which we have $\varphi=4\pi$ and $2\pi$ when $\bar{\Omega}/2\pi=0.32$ and $0.64$~MHz, with the gate fidelity error of about $8.0\times10^{-3}$ and $6.9\times10^{-3}$, respectively. Note that a correct estimate of $\varphi$ should include gap times between pulses that we have ignored for the sake of simplicity.

Second, one can choose three types of Rydberg states $|r_{c1}\rangle$, $|r_{c2}\rangle$, and $|r_{t}\rangle$ that are excited specifically from the first control qubit, the second control qubit, and the target qubit, respectively. We consider the condition where $|r_{c1}\rangle$ and $|r_{t}\rangle$ are Rydberg eigenstates, while $|r_{c2}\rangle$ is a superposition state constructed from two different Rydberg eigenstates; as shown in Ref.~\cite{Shi2017pra}, it is possible to choose appropriate superposition in $|r_{c2}\rangle$ so that the two-atom state $|r_{c1}r_{c2}\rangle$ has a zero C-six vdWI coefficient, thus experiences an exact zero vdWI, but the interactions of the states $|r_{c1}r_{t}\rangle$ and $|r_{c2}r_{t}\rangle$ are still normally significant. In this case, there will be no phase accumulation for the states $|r_{c1}r_{c2}0\rangle$ and $|r_{c1}r_{c2}1\rangle$ for the input states $|000\rangle$ and $|001\rangle$, respectively, while the necessary blockade interaction between the control and target qubits can still arise during pulses 2-4, thus the Deutsch gate can still be realized. As shown in Ref.~\cite{Shi2017pra}, strong microwave fields should be used to protect the superposition state $|r_{c2}\rangle$ against vdWI-induced decay.

The first method is simpler to implement, but the parameters to achieve $e^{i\varphi}=1$ may lead to a gate of a larger fidelity error compared to the optimal value, as discussed above; it is also sensitive to the position fluctuation of the qubits that results in fluctuation of the vdWI unless efficient cooling of the qubits is employed. While the second method has extra experimental complexity of using microwave fields for the stabilization of the superposition, it may allow the realization of the Deutsch gate with optimal gate fidelity. Assuming that the phase accumulation for $|000\rangle$ and $|001\rangle$ can be eliminated by at lease one of the two methods described above, we study the intrinsic fidelity error below.

\begin{table}
  \begin{tabular}{|c|c|}
    \hline   Input state  & The time $T_{\text{Ry}}$ of being in Rydberg states \\ \hline
  $ |000\rangle$ & 2$T_x $ \\ \hline 
  $ |001\rangle$ & 2$T_x $ \\ \hline 
  $ |010\rangle$ & $T_x $\\ \hline 
  $ |011\rangle$ & $T_x $\\ \hline  
  $ |100\rangle$ & $T_x $\\ \hline 
  $ |101\rangle$ & $T_x $\\ \hline 
  $ |110\rangle$ & $\frac{\pi}{\bar\Omega}\left[\frac{\Omega_1^2}{\bar\Omega^2} +\left|\frac{\Omega_2|\Omega_2^2 -3\Omega_1^2 |}{\bar\Omega^3}\right|^2 \right]+\frac{\pi}{\sqrt2\Omega_3}\frac{1}{2}$\\ \hline 
  $ |111\rangle$ & $\frac{\pi}{\bar\Omega}\left[\frac{\Omega_2^2}{\bar\Omega^2} +\left|\frac{\Omega_1(\Omega_1^2 -3\Omega_2^2)}{\bar\Omega^3}\right|^2 \right]+\frac{\pi}{\sqrt2\Omega_3}\frac{1}{2}$\\ \hline 
  \end{tabular}
  \caption{ Times for the atom to be in Rydberg states for different input states in the Deutsch gate protocol, where $T_x= \frac{\pi}{\Omega_0} + 2\cdot\frac{2\pi}{\bar\Omega}+ \frac{\sqrt2\pi}{\Omega_3}$. \label{table0}
  }
\end{table}
\begin{table}
  \begin{tabular}{|c|c|c|c|  }
    \hline   Input state  & Transition & $\Omega_x$~(pulses-2, 3) &$\Omega_x$~(pulse-4)  \\ \hline
  $ |000\rangle$ & $|rr0\rangle \rightarrow  |rr1\rangle $ & $\frac{i\Omega_1\Omega_2}{4V+V/32}$ & $\frac{-\Omega_3^2}{4V+V/32}$   \\ \hline 
  $ |001\rangle$ &  $|rr1\rangle \rightarrow  |rr0\rangle $ & $\frac{i\Omega_1\Omega_2}{4V+V/32}$ & $\frac{-\Omega_3^2}{4V+V/32}$   \\ \hline 
  $ |010\rangle$ & $|r10\rangle \rightarrow  |r11\rangle $ & $\frac{i\Omega_1\Omega_2}{2V}$ & $\frac{-\Omega_3^2}{2V}$   \\ \hline 
  $ |011\rangle$ &  $|r11\rangle \rightarrow  |r10\rangle $ & $\frac{i\Omega_1\Omega_2}{2V}$ & $\frac{-\Omega_3^2}{2V}$   \\ \hline 
  $ |100\rangle$ & $|1r0\rangle \rightarrow  |1r1\rangle $ & $\frac{i\Omega_1\Omega_2}{2V}$ & $\frac{-\Omega_3^2}{2V}$   \\ \hline 
  $ |101\rangle$ & $|1r1\rangle \rightarrow  |1r0\rangle $ & $\frac{i\Omega_1\Omega_2}{2V}$ & $\frac{-\Omega_3^2}{2V}$   \\ \hline 
  $ |110\rangle$ & - &  - &  -\\ \hline 
  $ |111\rangle$ & - &  - &  -\\ \hline 
  \end{tabular}
  \caption{ Two-photon transitions formed via the large blockade shift $V$ or $2V+V/64$. \label{table2}
  }
\end{table}

\section{Decay error}

The duration for each of the four input states to stay in Rydberg state is listed in Table~\ref{table0} for the Deutsch gate protocol. According to Ref.~\cite{Beterov2009}, the lifetime of the state $|r\rangle=|84p_{3/2},m_J=3/2,m_I=-1/2\rangle$ is about $\tau=1.59$~ms~(313~$\mu$s) for a temperature of 4.2~(300)~K. The decay-induced fidelity error of the Deutsch gate averaged over the four input states is
\begin{eqnarray}
  E_{\text{decay}} &=&\overline{T}_{\text{Ry}}/\tau,   \label{decayE}
\end{eqnarray}
where
\begin{eqnarray}
  \overline{T}_{\text{Ry}} &=& T_x +\frac{\pi}{4\bar\Omega}  +\frac{\pi}{8\sqrt2\Omega_3}.
\end{eqnarray}

\section{Blockade error}
Another fidelity error can be called the blockade error. During pulse-$1$ and pulse-$5$, the input state $|00\beta\rangle$ can not be fully converted back and forth to the state $|rr\beta\rangle$ because there is a residue blockade $V/64$ between the two control qubits when they are in Rydberg state. This contributes an error of about~\cite{Saffman2005}
\begin{eqnarray}
 E_{\text{bl}} &=&2(V/64)^2/\Omega_0^2
\end{eqnarray}
to the gate fidelity.

\section{Two-photon transition error }
During pulses 2-4, the blockade interaction $V$~(or $2V+V/64$) that can arise for the input states $|01\beta\rangle$ and $|10\beta\rangle$~(or $|00\beta\rangle$) is very large compared with $\{\Omega_1,\Omega_2,\Omega_3\}$, resulting in two-photon transitions~\cite{Shi2014} of effective Rabi frequencies $\Omega_x$ depending on the input states, as listed in Table~\ref{table2}. For pulses-2 and 3, the leakage transition has a duration of $2t=4\pi/\bar\Omega$~[see Eq.~(3)], thus the population loss is $\sin^2(2t\Omega_x/2)$ with $\Omega_x$ listed in Table~\ref{table2} for each input state. Similarly, for pulse-4, the time for the leakage transition is $t'=\sqrt2\pi/\Omega_3$, and the population loss is $\sin^2(t'\Omega_x/2)$. When neglecting $V/32$ in $4V+V/32$ of Table~\ref{table2}, the average error caused by such population loss is
\begin{eqnarray}
 E_{\text{2-ph}} &=&\frac{1}{4}\left[\sin^2\left(\frac{\Omega_1\Omega_2t}{4V} \right) +  \sin^2\left(\frac{\Omega_3^2t'}{8V} \right)  \right] \nonumber\\&& +\frac{1}{2}\left[\sin^2\left(\frac{\Omega_1\Omega_2t}{2V} \right) +  \sin^2\left(\frac{\Omega_3^2t'}{4V} \right)  \right] .
\end{eqnarray}

There is also an error caused by leakage to unwanted nearby levels. As analyzed in Ref.~\cite{Shi2017}, however, such error is much smaller than the blockade error, thus can be ignored. To briefly show why such leakage is negligible, we note that the two nearest $p_{3/2}$ Rydberg states around $|r\rangle$ that can be populated are $|85(83)p_{3/2},m_J=3/2,m_I=-1/2\rangle$, with detunings of about $\Delta=12.4~(-12.9)\times2\pi$~GHz, which is larger than the ground-state hyperfine splitting $\omega_g=9.19\times2\pi$~GHz. So, the leakage error during pulses-1 and 5~(or pulses 2-4) is proportional to $\Omega_0^2/(\Delta-\omega_g)^2$~[or $\Omega_k^2/(\Delta\pm\omega_g)^2$, where $k=1,2$ or $3$], which is on the order of $10^{-5}$~(or $10^{-7}$) for parameters in Fig.~4, thus can be neglected. In this case, the total gate fidelity error is given by
\begin{eqnarray}
 1-\mathcal{F}&=&E_{\text{decay}}+ E_{\text{bl}} + E_{\text{2-ph}},
\end{eqnarray}
based on which the rescaled fidelity error is shown in Fig.~4.

\end{document}